\documentclass[journal]{IEEEtran}
\usepackage{}
\usepackage{pifont}
\usepackage{times,amsmath,color,amssymb,graphicx,epsfig,cite,psfrag,subfigure}
\usepackage{mathrsfs}
\usepackage{amsfonts}
\usepackage{amsthm}
\usepackage{algorithm}
\usepackage{algpseudocode}

\long\def\symbolfootnote[#1]#2{\begingroup%
\def\thefootnote{\fnsymbol{footnote}}\footnote[#1]{#2}\endgroup}

\begin{document}
\title{Incentive Mechanisms  for Motivating Mobile Data Offloading in Heterogeneous Networks: A Salary-Plus-Bonus Approach\vspace{-1mm}}
\author{~~~~~Xin Kang,~\IEEEmembership{Member,~IEEE},  ~Sumei
Sun,~\IEEEmembership{Fellow,~IEEE}, ~Jing Yang,~\IEEEmembership{Member,~IEEE},
\thanks{This paper has been presented in \cite{0} at IEEE ICC. }
\thanks{X. Kang is with the National Key Laboratory of Science and Technology on Communications,  University of Electronic Science and Technology of China, Chengdu, China 611731. (E-mail: kang.xin@uestc.edu.cn). }
\thanks{S. Sun is with Institute for Infocomm Research, 1 Fusionopolis Way,
$\#$21-01 Connexis, South Tower, Singapore 138632 (E-mail: sunsm@i2r.a-star.edu.sg).}
\thanks{J. Yang is with Singapore PowerGrid Ltd, Singapore Power Group, 2 Kallang Sector, Singapore 349277. (E-mail: jing@singaporepower.com.sg). }} \maketitle

\begin{abstract}
In this paper, a \emph{salary-plus-bonus} incentive mechanism is proposed to motivate WiFi Access Points (APs) to provide data offloading service for mobile network operators (MNOs).  Under the proposed salary-plus-bonus scheme, WiFi APs are rewarded not only based on offloaded data volume but also based on the quality of their offloading service. The interactions between WiFi APs and the MNO under this incentive mechanism are then studied using Stackelberg game. By differentiating whether WiFi APs are of the same type (e.g. offloading cost and quality), two cases (\emph{homogeneous} and \emph{heterogeneous}) are studied.  For both cases, we derive the best response functions for WiFi APs (i.e. the optimal amount of data to offload), and show that the Nash Equilibrium (NE) always exists for the subgame. Then, given WiFi APs' strategies, we investigate the optimal strategy (i.e. the optimal salary and bonus) for the MNO to maximize its utility. Then, two simple incentive mechanisms, referred to as the \emph{salary-only} scheme and the  \emph{bonus-only} scheme, are presented and studied using Stackelberg game. For both of them, it is shown that the Stackelberg Equilibrium (SE) exists and is unique.  We also show that the salary-only scheme is more effective in offloading more data,  and the bonus-only scheme is more effective in selecting premium APs (i.e. providing high-quality offloading service at low cost), while the salary-plus-bonus  scheme can strike a well balance between the offloaded data volume and the offloading quality.  
\end{abstract}


\section{Introduction}
\subsection{Background and Motivation}
With the rapid development of smart phones and mobile broadband services, data
usage over the cellular network increases dramatically recently \cite{Cisco}. The unprecedented explosion of mobile data traffic poses new challenges to the current cellular networks. For example, in crowed areas such as metro areas and during peak hours, most 4G networks are overloaded\cite{Wavion}. The quality of experience in these overloaded
areas is therefore affected, e.g., low data transmission rate, access to some mobile applications, etc. Upgrading the cellular network to the more advanced 5G network \cite{XKangFullDuplex} or deploying more base
stations with smaller cell size \cite{XKangJSAC2011}  may be a viable solution for the aforementioned problem. However, these approaches may incur increase in infrastructure cost.

From the mobile operator's perspective, a more cost-effective approach is to offload some of the mobile traffic to existing WiFi networks, which is often referred to as \emph{WiFi offloading}. WiFi offloading is also a practical and readily available solution for a few reasons: (i) most of the mobile data services are created by smart phones which already have built-in WiFi modules, and (ii) WiFi's high data transmission rate. IEEE 802.11n WiFi can deliver data rates as high as 600Mbps and IEEE 802.11ac can deliver up to 6.933Gbps \cite{CiscoWiFi}, which is much faster than 4G. Recent research papers \cite{ABalasubramanian}--\kern-1mm\cite{CKHo} also demonstrated the feasibility and  effectiveness of WiFi offloading in relieving the data traffic burden of cellular networks. In \cite{ABalasubramanian}, the feasibility of augmenting 3G using WiFi was studied. In \cite{KLee}, performance of data offloading through WiFi networks for metropolitan areas was investigated. The numbers of WiFi APs needed for data offloading in large metropolitan areas was studied in \cite{SDimatteocy}.The load-balancing and
user-association problem for offloading in heterogeneous networks
were investigated in\cite{Jeff1}. In \cite{CKHo}, the authors investigated data offloading schemes for load coupled networks, and showed that the optimal loading is tractable when proportional fairness is considered.

Though WiFi offloading is a promising technology and has many advantages, without economic incentives, WiFi APs may be reluctant to provide data offloading service for the MNO. This is because providing offloading service for the MNO will inevitably incur additional operation cost, such as energy cost, data-usage cost. Besides, when providing data service for guest users from the cellular network, WiFi APs may have to sacrifice its own users' benefit, such as bandwidth, transmission rate, and quality of service. Therefore, there is a compelling need to design effective incentive mechanisms to motivate WiFi APs to participate in WiFi offloading.

\subsection{Related Work}
Incentive mechanisms to motivate WiFi APs to providing data offloading services or to motivate mobile users to offload data to WiFi APs have been studied in \cite{LGao}--\kern-1mm\cite{6}. In \cite{LGao}, the authors proposed the so-called market-based data offloading where the MNO pays WiFi APs for offloading traffic. An offloading game between the MNO and WiFi APs was formulated to study the pricing strategy of the MNO. In \cite{LGao-JSAC}, the authors considered  a one-to-many bargaining game among the MNO and APs, and analyzed the bargaining solution under the sequential bargaining and the concurrent bargaining, respectively. In \cite{7}, a three-stage  game was formulated to study the data offloading with price-taking and price-setting APs. In \cite{Xkang}, the authors investigated optimal user association strategies for a HetNet where the MNO pays third-party WiFi APs for providing data offloading service. However, in \cite{LGao, LGao-JSAC,7,Xkang}, the MNO pays WiFi APs only based on the offloaded data volume, while the quality of data offloading service was considered in  the incentive mechanism. 

In \cite{JLeeSDP}, the authors focused on the interactions between the MNO and  mobile users. The MNO rewards mobile users if they direct their delay-tolerant data service to WiFi APs. The economic benefits brought to the MNO and users due to the delayed WiFi offloading were then studied. In \cite{5}, the authors  investigated the tradeoff between the amount of traffic being offloaded and the users' satisfaction. An incentive framework to motivate users to leverage their delay tolerance for cellular traffic offloading was proposed. In \cite{6}, the authors studied the load-balancing problem for data offloading, and designed a  quality-price contract to motivate users to make proper association strategy. However, the proposed incentive mechanisms in \cite{JLeeSDP,5,6} are aimed at providing incentives for mobile users rather than WiFi APs.

There are also works \cite{8,9,10,11,12} investigating mobile data  offloading from other perspectives.  In \cite{8}, the authors studied optimal scheduling for incentivizing   WiFi offloading under the energy constraint. 
A secrecy-based energy-efficient data offloading with dual connectivity was studied in \cite{9}.  In \cite{10}, an energy-aware data offloading scheme via device-to-device cooperations was proposed.  In \cite{11}, the authors showed that WiFi data offloading achieves better performance than resource sharing when the number of WiFi users is below a threshold. In \cite{12}, a reverse data offloading scheme to offload WiFi data to  LTE-U (LTE in unlicensed band) was studied. 

\subsection{Main Contribution}
In this paper, we  consider  a heterogeneous network with a MNO and multiple third-parry WiFi APs.  Each WiFi has its home users (HUs) to serve, and it also has certain leftover bandwidth for providing data offloading service for the MNO.  We design  incentive mechanisms to motivate these WiFi APs to provide data offloading for the MNO. The main contribution is summarized as follows.
\begin{itemize}
	\item We propose a \emph{salary-plus-bonus} reward scheme as an incentive to motivate WiFi APs for providing data offloading service for the MNO. Particularly, the proposed incentive mechanism rewards WiFi APs not only based on the amount of data offloaded but also  the quality of their offloading service. 
	\item Under the proposed \emph{salary-plus-bonus} scheme, we investigate the interactions between the MNO and WiFi APs using Stackelberg game. We derive the best response functions for WiFi APs which lead to the subgame Nash Equilibrium (NE). We also investigate the optimal bonus and salary rate that the MNO should set in order to maximize its utility.
	\item We study the formulated Stackelberg game under two different scenarios: \emph{Homogeneous} APs and \emph{Heterogeneous}  APs. For the  \emph{Homogeneous}  case, all WiFi APs are assumed to be the same type (e.g. offloading quality and offloading cost). Both the subgame NE and the MNO's optimal strategy are obtained in  closed-form solutions. For the \emph{heterogeneous } case, the subgame NE is obtained in closed-form for the two-AP case.  For  the multi-AP case, we show that the subgame NE can be found by the simplicial method, and the MNO's optimal strategy can be found by a two-dimension grid search. 
	\item For the purpose of comparison, we also investigate two simplified versions of the \emph{salary-plus-bonus} scheme: the \emph{salary-only} scheme and the \emph{bonus-only} scheme. For both schemes, we develop high-efficiency and low-complexity algorithms to find the optimal strategy for both WiFi APs and the MNO. It is shown that the salary-only scheme is more effective in motivating more APs to offload more data, while the bonus-only scheme is more effective in selecting premium APs which can provide high-quality offloading service at low cost. 
	\item We investigate the performance of the aforementioned  incentive mechanisms by numerical simulations. To study the performance of the proposed \emph{salary-plus-bonus} scheme for heterogeneous networks with  large-size, we develop a  low-complexity suboptimal algorithm to quickly find the strategy of APs' and the MNO.  It is shown that the proposed \emph{salary-plus-bonus} scheme can strike a good balance between the offloading quality and  offloaded data volume. 
\end{itemize}

\subsection{Organization of this Paper}
The organization of the rest of this paper is as follows. Section \ref{sec-systemmodel} presents the system model and the Stackelberg game formulation for the proposed salary-plus-bonus scheme. In Section \ref{sec-optimalHetAP} and \ref{sec-homoAPs}, we study the optimal offloading strategy of WiFi APs and the optimal strategy of the MNO for the homogeneous and the heterogeneous case, respectively. In Section \ref{sec-salaryonly} an \ref{sec-bonusonly}, we  present the Stackelberg game formulations for the salary-only and the bonus-only scheme, respectively. We also derive the Stackelberg equilibrium (i.e.,the optimal solutons for APs and  the MNO) for  the two cases, respectively. In Section \ref{sec-numbericalresults}, the numerical results are presented to compare the performance of the three proposed incentive mechanisms. Especially, we develop a suboptimal algorithm for the salary-plus-bonus scheme to efficiently find the strategy of APs and the MNO for heterogeneous networks with large size. Finally, Section \ref{sec-conclusion} concludes the paper.  

\section{System Model and Game Formulation}\label{sec-systemmodel}
\begin{figure}[t]
        \centering
        \includegraphics*[width=8cm]{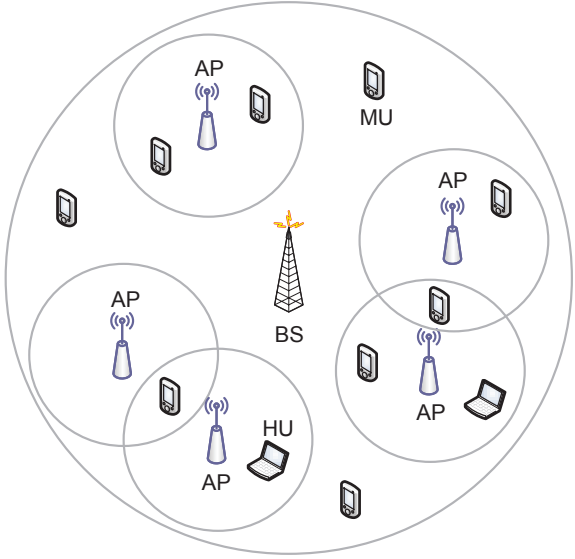}
        \caption{Data offloading in HetNet with third-party WiFi APs}
        \label{Fig-OffloadingIncenFig1}
\end{figure}

A Stackelberg game \cite{XKangJSAC2011,XkangTMC} is a strategic game that consists
of a leader and several followers competing with each
other on certain resources. The leader moves first and the
followers move subsequently. In this paper, as shown in Fig. \ref{Fig-OffloadingIncenFig1}, we consider a heterogeneous network with a MNO and multiple WiFi APs. The set of WiFi APs is denoted by $\mathcal{N}$. All the WiFi APs can provide data offloading service for the MNO. In particular, we consider the case that each WiFi AP may have its home users (HUs) and thus it should reserve certain bandwidth for its HUs. In this paper, like the existing work \cite{LGao}--\kern-1mm\cite{5}, we investigate the data offloading problem and design incentive mechanism purely from the data level. Thus, the related physical layer and MAC layer issues to implement the data offloading schemes are out of concern of this paper. In this paper, we formulate
the MNO as the leader, and the WiFi APs as the followers.
The MNO (leader) announces a salary and a bonus to the WiFi APs.
Then, each WiFi AP (follower) determines its optimal
amount of data (that it intends to offload) to maximize its utility
based on the salary and the bonus. Thus, the Stackelberg
Game consists of two parts: the game at the WiFi APs and the game at the MNO, which are introduced in the following two subsections, respectively.

\subsection{The Game at WiFi APs}
Let $p$ denote the pay rate, i.e., cash paid to a WiFi AP on per unit of data offloaded.
Let $B$ denote the total amount of bonus paid to all WiFi APs for data offloading. Then, the utility function of an arbitrary WiFi AP can be modelled as
\begin{align}\label{eq-APUtility}
\mathcal {U}_i^F(\boldsymbol{d},p,B)=\mathcal {S}_i(p, d_i)+\mathcal {B}_i(B, \boldsymbol{w}, \boldsymbol{d})-\mathcal {C}_i(d_i),
\end{align}
where $\boldsymbol{d}\triangleq[d_1, \cdots, d_N]^T$ with entry $d_i$ denoting the amount of data that $\mbox{AP}_i$ offloads for the MNO, and $\boldsymbol{w}\triangleq[w_1, \cdots, w_N]^T$ with entry $w_i$ denoting the quality of offloading service provided by $\mbox{AP}_i$.

It is observed from \eqref{eq-APUtility} that each AP's utility function consists of three parts: $\mathcal {S}_i(p, d_i)$, $\mathcal {B}_i(B, \boldsymbol{w}, \boldsymbol{d})$, and $\mathcal {C}_i(d_i)$. In the following, we present how to model them under the proposed incentive mechanism.

 \emph{Salary}: $\mathcal {S}_i(p,d_i)$ denotes the salary of $\mbox{AP}_i$, i.e. the payment received for providing data offloading service
  for the MNO. $\mathcal {S}_i(p,d_i)$ is a function of $d_i$ and
  $p$. Intuitively, the more work you have done, the more payment you
  should receive. Thus,  $\mathcal
  {S}_i(p,d_i)$ should be an increasing function of $d_i$. Besides, $\mathcal  {S}_i(p,d_i)$ also should be an increasing function of
$p$, since the higher the pay rate is, the more payment you will
receive. In this paper, for simplicity, we use a linear function to model the salary,
which is given as
\begin{align}
\mathcal {S}_i(p,d_i)=pd_i.
\end{align}

\emph{Bonus}: $\mathcal {B}_i(B,
\boldsymbol{w}, \boldsymbol{d})$ denotes the bonus
paid to $\mbox{AP}_i$ by the MNO. 
In game theory literature, there are many bonus distribution models, such as equal share, Shapley value \cite{GameTheory1993}, marginal contribution.
In this work, to better stimulate WiFi APs' to participate in data offloading, we use the weighted proportional share model, which is
\begin{align}\label{eq-BounsAPi}
\mathcal {B}_i(B, \boldsymbol{w},
\boldsymbol{d})=\frac{w_i d_i}{\sum_{j \in \mathcal {N}}
w_jd_j}B.
\end{align}
It is observed from \eqref{eq-APUtility}
that $\mbox{AP}_i$'s bonus $\mathcal {B}_i$ not only depends on
its own performance (quality of offloading service $w_i$ and amount of data offloaded $d_i$) but also depends on other AP's
performance $(\boldsymbol{w}_{-i}$ and $\boldsymbol{d}_{-i})$. This is analogous to the bonus system in human's society, where a staff's bonus not only depends on his/her own performance but also depends on other staffs' performance.

\emph{Cost}: $\mathcal {C}_i(d_i)$ denotes the cost incurred
  when $\mbox{AP}_i$ provides data offloading service for the MNO.
  Usually, when a WiFi AP provides data service for more users, it will incur more cost, such as electricity cost, data usage cost, and etc. In general, the cost increases with the increasing of the amount of data
  offloaded. Thus, in this work, we model the cost as
\begin{align}
\mathcal {C}_i(d_i)=c_id_i,
\end{align}
where $c_i$ is a positive constant that relates the amount of data offloaded to the cost of $\mbox{AP}_i$.

Under the Stackelberg game formulation, the amount of data that
$\mbox{AP}_i$ intends to offload depends on the pay rate $p$ and the
bonus $B$. In general, if the MNO sets a high $p$ and a high $B$, $\mbox{AP}_i$ is willing to offload more data, and vice versa. Thus, each WiFi AP has to determine
its optimal $d_i^*$ given $p$, $B$ and other APs' offloading amount. Mathematically, the problem can be written as

\underline{\textbf{Problem 1:}}
\begin{align}
\max_{d_i\ge0}~~&pd_i+\frac{w_i d_i}{\sum_{j \in \mathcal {N}}
w_jd_j}B-c_id_i,\\
\mbox{s.t.}~~&d_i\le T_i,\label{eq-con}
\end{align}
where $T_i\triangleq T^C_i-T_i^H$. $T^C_i$ is the maximum amount of data can be admitted within the $\mbox{AP}_i$'s capacity and $T_i^H$ is the data quota reserved for its HUs. Thus, the constraint \eqref{eq-con} represents the maximum amount of data that a WiFi AP can offload.

\subsection{The Game at the MNO}
In this subsection, we define the MNO's utility and present the game at the MNO. Without loss of generality, in this paper, we define the MNO's utility function as
\begin{align}
\mathcal {U}^L(p, B, \boldsymbol{d})\triangleq\mathcal {R}^L(p, B, \boldsymbol{d})-\mathcal
{C}^L(p, B,\boldsymbol{d}),
\end{align}
where $\mathcal {R}^L(p, B, \boldsymbol{d})$ is the
payoff/benefit gained from offloading data, $\mathcal
{C}^L(p, B, \boldsymbol{d})$ is the cost incurred due to data offloading.

Note that the MNO's utility function consists of two parts: payoff and cost. Both of them are functions of $p$, $B$, and $\boldsymbol{d}$. In the following, we present how to model them under the proposed incentive mechanism.

\emph{Payoff:} The MNO's payoff is the benefit or reward gained from offloading data.  In this paper, we model the MNO's payoff as
\begin{align}
\mathcal {R}^L(p, B, \boldsymbol{d})=\lambda f(\boldsymbol{d}),
\end{align}
where $f(\boldsymbol{d})$ is the \emph{offloading gain} for the MNO, and $\lambda$ is a positive constant known as \emph{offloading gain coefficient } converting the offloading gain into monetary reward. In this paper, we use a log function to model the offloading gain, i.e.,
\begin{align}\label{eq-OBE}
f(\boldsymbol{d})\triangleq \log_2\left(1+\sum_{i \in \mathcal
{N}}d_i\right).
\end{align}
Though other functions (such as linear functions or exponential functions) can also be used to model the offloading gain, log functions are shown in literatures to be more
suitable to representing the relationship between the network performance and a large class of elastic data traffics \cite{FPKelly}.
It is observed from \eqref{eq-OBE} that when
the amount of data offloaded is zero ($\sum_{i \in \mathcal
{N}}d_i=0$), the offloading gain $f$ is
zero. Besides, the offloading gain increases with
the increasing of the amount of data offloaded. These indicate that \eqref{eq-OBE} is able to capture the relationship between the MNO's benefit and the data offloaded.

\emph{Cost:} The MNO's cost includes two parts: the salary and the bonus. With a unified pay rate $p$, the total salary paid to WiFi APs is $p\sum_{i \in \mathcal {N}}d_i$. We assume that the bonus that the MNO intends to hand out is $B$. Thus, the cost function of the MNO can be modelled as
      \begin{align}
\mathcal
{C}^L(p,B,\boldsymbol{d})=p\sum_{i \in \mathcal {N}}d_i+B.
      \end{align}

As pointed out in the previous subsection, the amount of data that each AP intends to offload depends on the pay rate $p$ and the
bonus $B$. Thus, the MNO can easily control the
total amount of data offloaded to WiFi APs by controlling $p$ and $B$. However, the benefit of the MNO received from data offloading also
depends on $p$ and $B$. Setting high pay rate and high bonus can
help the MNO offload more data, however, this also increases the
operating cost of the MNO. Therefore, the MNO needs to find the optimal
$p^*$ and $B^*$ in order to maximize its utility. Mathematically, the problem can be written as

\underline{\textbf{Problem 2:}}
\begin{align}
\max_{p,~ B}~~&\lambda\ln \left(1+\sum_{i \in \mathcal
{N}}d_i(p,B)\right)-p\sum_{i \in \mathcal {N}}d_i(p,B)-B,\\
\mbox{s.t.}~~&~p\ge 0,~B\ge 0.
\end{align}

\subsection{Stackelberg Equilibrium and Subgame Nash Equilibrium}
Problem 1 and Problem 2 together
form a Stackelberg game. The objective of this game is to find
the Stackelberg Equilibrium (SE) point(s) from which neither
the MNO (leader) nor the WiFi APs (followers) have incentives to deviate. For the
proposed Stackelberg game, the SE is defined as follows.

\underline{\textbf{Definition 1 (Stackelberg Equilibrium):}} Let $d_i^*$ be the solution for Problem 1 and
$(p^*,B^*)$ be the solution for Problem 2. Then, the point $(\boldsymbol{d}^*,p^*,B^*)$ is a SE for the proposed Stackelberg game if for any $(\boldsymbol{d},p,B)$,
the following conditions are satisfied:
\begin{align}
\mathcal {U}^L(p^*, B^*, \boldsymbol{d}^*) &\ge \mathcal {U}^L(p, B, \boldsymbol{d}^*),\\
\mathcal {U}_i^F(d_i^*,p^*,B^*) &\ge \mathcal {U}_i^F(d_i,p^*,B^*),\forall i.
\end{align}
where $\mathcal {U}^L$ and $\mathcal {U}^F_i$  are the utilities of the MNO and the WiFi $\mbox{AP}_i$, respectively.

In the proposed game, it is not difficult to see that WiFi APs strictly compete in a non-cooperative fashion. Therefore, a non-cooperative subgame is formulated at WiFi APs' side. For a non-cooperative game, NE is defined as the operating point(s) at which no player can improve utility by changing its strategy unilaterally, assuming everyone else continues to use its current strategy. Mathematically, it is defined as follows.

\underline{\textbf{Definition 2 (Nash Equilibrium):}} Let $(d_i^*, \boldsymbol{d}_{-i}^*)$ be the solution for Problem 1. Then, the point $(d_i^*, \boldsymbol{d}_{-i}^*)$ is a NE for the non-cooperative subgame if for any $(d_i, \boldsymbol{d}_{-i}^*)$,
the following conditions are satisfied:
\begin{align}
\mathcal {U}_i^F(d_i^*, \boldsymbol{d}_{-i}^*) &\ge \mathcal {U}_i^F(d_i,\boldsymbol{d}_{-i}^*),\forall i.
\end{align}

For the Stackelberg game formulated here, the SE  can be obtained by finding its subgame NE first. Then, given the subgame NE, the best response of the MNO can be readily obtained by solving Problem 2. In the following section, we investigate the optimal solution and analyze the equilibrium for the formulated data offloading game.  

\section{Optimal Solutions for Heterogeneous APs} \label{sec-optimalHetAP}
To find the SE, the optimal strategies for the followers
(WiFi APs) must be obtained first, and then the leader
(MNO) derives its optimal strategy on those of
the followers. This is also known as \emph{backward induction} in game-theoretic studies \cite{XKangJSAC2011,XkangTMC}. Using this method, the optimal strategies for the formulated game are derived in the following two subsections.
\subsection{Optimal Strategies of WiFi APs}
To find the optimal strategies of WiFi APs, we first look at the best response of each WiFi AP given $p$, $B$ and other APs' strategies, which is given in the following theorem.

\underline{\textbf{Theorem 1}:} The best response function of $\mbox{AP}_i$ is
\begin{align}\label{eq-theo1}
d_i^*=\left\{\kern-1mm\begin{array}{ll}
             0, &\mbox{if}~a_i>0~\mbox{and}~B\le a_i\frac{z_i}{w_i}, \\
             T_i, &\mbox{if}~a_i\le0~\mbox{or}~B\ge
             a_i\frac{z_i}{w_i}(1+\frac{w_iT_i}{z_i})^2,\\
\sqrt{\frac{Bz_i}{a_iw_i}}-\frac{z_i}{w_i},&\mbox{otherwise}.
           \end{array}
\right.
\end{align}
where \begin{align}z_i\triangleq\sum_{j \in \mathcal {N}/\{i\}} w_jd_j~\mbox{and}~a_i\triangleq{c_i-p}.\end{align}

\begin{proof}
Take the derivative of $\mathcal {U}_i^F$ with respect to $ d_i$, we
have
\begin{align}
\frac{\partial\mathcal {U}_i^F}{\partial d_i}=p+\frac{Bw_i\sum_{j
\in \mathcal {N}/\{i\}} w_jd_j}{\left(\sum_{j \in \mathcal {N}}
w_jd_j\right)^2}-c_i,
\end{align}
\begin{itemize}
  \item Case 1: $p-c_i\ge 0$. When $p-c_i\ge 0$, $\frac{\partial\mathcal {U}_i^F}{\partial d_i}$ is
always positive, which indicates $\mathcal {U}_i^F$ is monotonically
increasing with $d_i$.  Thus, $\mathcal {U}_i^F$ attains its maximum
when $d_i^*=T_i$.
  \item Case 2: $p-c_i< 0$. When $p-c_i< 0$, let $\frac{\partial\mathcal {U}_i^F}{\partial
  d_i}\big|_{d_i=d_i^\circ}=0$, we have
  \begin{align}
  d_i^\circ=\sqrt{\frac{B\sum_{j \in \mathcal {N}/\{i\}}
  w_jd_j}{w_i(c_i-p)}}-\frac{\sum_{j \in \mathcal {N}/\{i\}}
  w_jd_j}{w_i}.
  \end{align}

Since$ \frac{\partial^2\mathcal {U}_i^F}{\partial
d_i^2}=-\frac{2Bw_i^2\sum_{j \in \mathcal {N}/\{i\}}
w_jd_j}{\left(\sum_{j \in \mathcal {N}} w_jd_j\right)^3}<0$ ,
$\mathcal {U}_i^F$ is concave in $d_i$, and it follows that
\begin{align}
d_i^*=\left\{\kern-1mm\begin{array}{ll}
             0, &~\mbox{if}~d_i^\circ \le 0, \\
             d_i^\circ, &~\mbox{if}~0<d_i^\circ<T_i,\\
             T_i,&~\mbox{if}~d_i^\circ\ge T_i.
           \end{array}
\right.
\end{align}
\end{itemize}
Then, let $z_i\triangleq\sum_{j \in \mathcal {N}/\{i\}} w_jd_j$ and
$a_i\triangleq{c_i-p}$. Theorem 1 follows by combining the results
obtained in Case 1 and 2.\end{proof}

Now, we investigate the NE of this subgame.  It is observed from Problem 1 that APs' strategy set is compact and convex. The utility of $\mbox{AP}_i$ is continuous and concave in $d_i$, and continuous in $\boldsymbol{d}_{-i}$. Thus, according to the
\textbf{debreu-glicksberg-fan} theorem \cite{GameTheory1993}, a pure strategy NE exists. However, the subgame NE can not be obtained in closed-form due to the high complexity. Numerically, the subgame NE can be computed by the simplicial method \cite{PHerings}. The basic idea is to solve the non-linear equilibrium problem by solving a piecewise linear approximation of the problem.  For the purpose of illustration, we show the results for the two-AP case here.  

Assume $c_2>c_1$,  the subgame NE denoted by $(d_1^{ne}, d_2^{ne})$  can be obtained by case-by-case discussion, which is given as

\textit{Case I. } When $p\ge c_2>c_1$, the NE is
\begin{align}
(d_1^{ne}, d_2^{ne})=(T_1, T_2).
\end{align}

\textit{Case II. }When $c_2>p\ge c_1$, the NE is
\begin{align}
&~(d_1^{ne}, d_2^{ne})=\nonumber\\
&\left\{\kern-1.5mm
\begin{array}{ll}
(T_1, 0), & \mbox{if}~0\le B< \frac{a_2}{w_2}w_1T_1,\\
\left(T_1, \sqrt{\frac{Bw_1T_1}{w_2a_2}}\kern-1mm-\kern-1mm\frac{w_1T_1}{w_2}\right) ,& \mbox{if}~\frac{a_2w_1T_1}{w_2}\kern-0.5mm\le\kern-0.5mm B\kern-1mm<\kern-1mm\frac{a_2}{T_1}\frac{(w_1T_1+w_2T_2)^2}{w_1w_2}, \\
\left(T_1, T_2\right) ,& \mbox{if}~\frac{a_2}{T_1}\frac{(w_1T_1+w_2T_2)^2}{w_1w_2}\le B.
\end{array}
\right.
\end{align}

\textit{Case III.}  When $c_2>c_1>p\ge 0$ and $a_1T_1<a_2T_2$, the NE is
\begin{align}
&(d_1^{ne},d_2^{ne})=\nonumber\\
&\left\{\kern-1.5mm
\begin{array}{ll}
\left(\frac{Bw_1w_2a_2}{(w_1a_2+w_2a_1)^2}, \frac{Bw_1w_2a_1}{(w_1a_2+w_2a_1)^2}\right)& \mbox{if}~B\in\mathcal {A}_1,\\
\left(T_1, \sqrt{\frac{Bw_1T_1}{w_2a_2}}-\frac{w_1T_1}{w_2}\right),& \mbox{if}~B\in \mathcal {A}_2,\\
\left(T_1, T_2\right),& \mbox{if}~B\in \mathcal {A}_3.
\end{array}
\right.
\end{align}
where the regions are defined as: $\mathcal {A}_1\triangleq \left(0,\frac{(w_1a_2+w_2a_1)^2}{w_1w_2}\frac{T_1}{a_2}\right)$, $\mathcal {A}_2\triangleq \left[\frac{(w_1a_2+w_2a_1)^2}{w_1w_2}\frac{T_1}{a_2},\frac{a_2}{T_1}\frac{(w_1T_1+w_2T_2)^2}{w_1w_2}\right) $, and $\mathcal {A}_3\triangleq \left[\frac{a_2}{T_1}\frac{(w_1T_1+w_2T_2)^2}{w_1w_2},\infty\right)$.

\textit{Case IV.} When $c_2>c_1>p\ge 0$ and $a_1T_1>a_2T_2$, the NE is
\begin{align}
&(d_1^{ne},d_2^{ne})=\nonumber\\
&\left\{\kern-1.5mm
\begin{array}{ll}
\left(\frac{Bw_1w_2a_2}{(w_1a_2+w_2a_1)^2}, \frac{Bw_1w_2a_1}{(w_1a_2+w_2a_1)^2}\right)& \mbox{if}~B\in\tilde{\mathcal {A}}_1,\\
\left(\sqrt{\frac{Bw_2T_2}{w_1a_1}}-\frac{w_2T_2}{w_1},T_2\right),& \mbox{if}~B\in \tilde{\mathcal {A}}_2,\\
\left(T_1, T_2\right),& \mbox{if}~B\in \tilde{\mathcal {A}}_3.
\end{array}
\right.
\end{align}
where the regions are defined as: $\tilde{\mathcal {A}}_1\triangleq \left(0,\frac{(w_1a_2+w_2a_1)^2}{w_1w_2}\frac{T_2}{a_1}\right)$, $\tilde{\mathcal {A}}_2\triangleq \left[\frac{(w_1a_2+w_2a_1)^2}{w_1w_2}\frac{T_2}{a_1},\frac{a_1}{T_2}\frac{(w_1T_1+w_2T_2)^2}{w_1w_2}\right) $, and $\tilde{\mathcal {A}}_3\triangleq \left[\frac{a_1}{T_2}\frac{(w_1T_1+w_2T_2)^2}{w_1w_2},\infty\right)$.

From the above results, we can observe that:
\begin{itemize}
	\item For any given $p$ and $B$,  the NE is unique.
		\item Different values of $p$ and $B$ result in different NE. This indicates that the NE is affected by both $p$ and $B$.
	\item The order of $a_iT_i$ has an impact on the NE. The WiFi AP with lower $a_iT_i$ is more likely to reach its capacity limit $T_i$ first.
		\item All APs will offload at their capacity limits when $p\ge\mbox{argmax}_i c_i$. 
\end{itemize}

\subsection{The Optimal Strategy of the MNO}
Now, given WiFi AP's strategies, we derive the optimal strategy of the MNO. For the MNO, the optimal strategy can not be obtained in closed-form since there is no explicit expression of WiFi APs' strategies.  Besides, given the subgame NE, Problem 2 is not a convex optimization problem. Thus, convex optimization techniques or existing convex optimizers can not be applied here. 

\subsubsection{Two-AP Case} For the two-AP case,  it is observed from its subgame NE that: (i) The optimal $p^*$ is bounded by $c_2$; (ii) For a given $p$, Problem 2 is concave in $B$ for each separate region of $B$ (such as $\mathcal {A}_1$, $\mathcal {A}_2$, and $\mathcal {A}_3$). Thus, the optimal $B$ for each separate region can be easily obtained using the convex optimization techniques. Then, the optimal $B^*$ can be obtained by comparing the maximum utility function of each separate region.  Thus, the optimal strategy of the MNO for the two-AP case can be obtained by the following two steps: 

Step I. For a given $p$, compute the optimal $B^*$. 

Step II. Search for the optimal $p^*$ over the region $[0, c_2]$.

\subsubsection{Multi-AP Case} For the multi-AP case, similar as the two-AP case, we can show that $p^*$ is bounded by $\mbox{argmax}_i c_i$. This is due to the fact that when $p=\mbox{argmax}_i c_i$, all APs will offload at their capacity limits . Thus, using a $p$ larger than $\mbox{argmax}_i c_i$ will not increase the MNO's payoff, but will increase the MNO's cost. Thus, it is concluded that $p^*$ is bounded by $\mbox{argmax}_i c_i$.  

Next, we can further show that $B^*$ is bounded by $\mbox{argmax}_i \frac{a_i}{w_i\sum_{j\in \mathcal{N}/i} w_jT_j}\left(\sum_{j\in \mathcal{N}} w_jT_j\right)^2$. The proof is as follows. It is observed from \eqref{eq-theo1} that $d_i^*=T_i$ when $B\ge
a_i\frac{z_i}{w_i}(1+\frac{w_iT_i}{z_i})^2$. Thus, if  $B=\mbox{argmax}_i \frac{a_i}{w_i\sum_{j\in \mathcal{N}/i} w_jT_j}\left(\sum_{j\in \mathcal{N}} w_jT_j\right)^2$, all APs will offload at their capacity limits. Thus, using a larger $B$  will not increase the MNO's payoff, but will increase the MNO's cost. Thus, it is concluded that $p^*$ is bounded by $\mbox{argmax}_i \frac{a_i}{w_i\sum_{j\in \mathcal{N}/i} w_jT_j}\left(\sum_{j\in \mathcal{N}} w_jT_j\right)^2$.

Since both $p^*$ and $B^*$ are bounded, the MNO' optimal strategy can be obtained by performing a two-dimension grid search over $p$ and $B$.

\section{Optimal Solutions for Homogeneous APs}\label{sec-homoAPs}
In this section, to obtain closed-form solutions and get useful insights, we assume all the WiFi APs are of the same type (homogeneous), i.e., $w_i=w,$ $c_i=c,$ $T_i=T,$ $\forall i$. 

\subsection{Optimal Strategies of WiFi APs}
Based on the best response function given in \eqref{eq-theo1}, by setting $w_i=w,$ $c_i=c,$ $T_i=T,$ $\forall i$, the NE can be easily computed as follows.
\begin{itemize}
	\item When $a\le0$, the NE is \begin{align}\label{eq-HomoNE1}d_i^{ne}=T,~ \forall i.\end{align}
	\item When $a>0$, the NE is \begin{align}\label{Eq-HomoNE}
	d_i^{ne}=\left\{\kern-1mm\begin{array}{ll}
	\frac{B(|\mathcal{N}|-1)}{a|\mathcal{N}|^2}, &\mbox{if}~\frac{B}{a}<\frac{|\mathcal{N}|^2T}{|\mathcal{N}|-1},\\
	T,&\mbox{if}~\frac{B}{a}\ge \frac{|\mathcal{N}|^2T}{|\mathcal{N}|-1},
	\end{array}
	\right.
	\end{align}
	where $|\cdot|$ denotes the cardinality of a set.
\end{itemize}


It is observed that (i) different values of $p$ and $B$ will result in different NE;
(ii) for given $p$ and $B$, the NE is unique; (iii) all the WiFi APs have the same strategy at the NE.

\subsection{The Optimal Strategy of the MNO}\label{Sec-HomoOpMNO}
Given APs' strategies, we now study the optimal strategy of the MNO. To find the optimal strategy of the MNO, we need to substitute the subgame NE given in \eqref{Eq-HomoNE} into Problem 2.

First, we look at the case that $a\le 0$, i.e., $p\ge c$.  For this case, the subgame NE is given by \eqref{eq-HomoNE1}. Substitute \eqref{eq-HomoNE1} into Problem 2, the MNO's utility maximization problem becomes
\begin{align}
\max_{p, B}~~&\lambda\ln\left(1+|\mathcal{N}|T\right)-p|\mathcal{N}|T-B,\\
\mbox{s.t.}~~&~p\ge 0,~B\ge 0.
\end{align}
The optimal solution of this problem is
\begin{align}\label{Eq-HomoMNO1}
p^*=c,~ B^*=0.
\end{align}
This results indicates that $p$ is bounded by $c$, and the MNO will never set a $p^*$ larger than $c$.

Now, we look at the case that $a>0$, i.e., $p<c$. For this case, we first present the following theorem.

\underline{\textbf{Theorem 2}:} For any given $p$ with $0\le p<c$, the best strategy of the MNO is
\begin{align}
B^*=\left\{\kern-1mm\begin{array}{ll}\label{eq-optimalB}
\frac{\lambda}{1+p\frac{|\mathcal{N}|-1}{a|\mathcal{N}|}} -\frac{1}{\frac{|\mathcal{N}|-1}{a|\mathcal{N}|}}, &\mbox{if}~\tilde{\mathcal {U}}^L< \hat{\mathcal {U}}^L,\\
\frac{a|\mathcal{N}|^2T}{|\mathcal{N}|-1},&\mbox{if}~\tilde{\mathcal {U}}^L\ge \hat{\mathcal {U}}^L.
\end{array}
\right.
\end{align}
where $\tilde{\mathcal {U}}^L\triangleq\lambda\ln\left(1+|\mathcal{N}|T\right)-p|\mathcal{N}|T-\frac{a|\mathcal{N}|^2T}{|\mathcal{N}|-1}
$ and $\hat{\mathcal {U}}^L\triangleq\ln \left(\frac{\lambda}{p+\frac{a|\mathcal{N}|}{|\mathcal{N}|-1}}\right)+\frac{a|\mathcal{N}|}{|\mathcal{N}|-1}+p-\lambda$.

\begin{proof}
	(i). When $B<\frac{a|\mathcal{N}|^2T}{|\mathcal{N}|-1}$, the MNO's utility can be written as
	\begin{align}\label{eq-UleaderGivenB}
	\max_{B\ge0} ~\lambda\ln \left(1+\frac{B(|\mathcal{N}|-1)}{a|\mathcal{N}|}\right)-p\frac{B(|\mathcal{N}|-1)}{a|\mathcal{N}|}-B,
	\end{align}
	It is easy to verify \eqref{eq-UleaderGivenB} is concave in $B$ by looking at its second-order derivative. Then, the optimal $\hat{B}^*$ can be obtained by setting the first-order derivative of \eqref{eq-UleaderGivenB} to zero, which is
	\begin{align}
	\lambda\frac{\frac{(|\mathcal{N}|-1)}{a|\mathcal{N}|}}{\frac{(|\mathcal{N}|-1)}{a|\mathcal{N}|}\hat{B}^*+1}-p\frac{(|\mathcal{N}|-1)}{a|\mathcal{N}|}-1=0,
	\end{align}
	Then, it follows that
	\begin{align}
	\hat{B}^*=\frac{\lambda}{1+p\frac{|\mathcal{N}|-1}{a|\mathcal{N}|}} -\frac{1}{\frac{|\mathcal{N}|-1}{a|\mathcal{N}|}}.
	\end{align}
	and the resultant utility is
	\begin{align}
	\hat{\mathcal {U}}^L=\ln \left(\frac{\lambda}{p+\frac{a|\mathcal{N}|}{|\mathcal{N}|-1}}\right)+\frac{a|\mathcal{N}|}{|\mathcal{N}|-1}+p-\lambda.
	\end{align}
	
	(ii). When $B\ge \frac{a|\mathcal{N}|^2T}{|\mathcal{N}|-1}$, it is easy to observe that
	\begin{align}
	\tilde{B}^*=\frac{a|\mathcal{N}|^2T}{|\mathcal{N}|-1},
	\end{align}
	and the resultant utility is
	\begin{align} \tilde{\mathcal {U}}^L=\lambda\ln\left(1+|\mathcal{N}|T\right)-p|\mathcal{N}|T-\frac{a|\mathcal{N}|^2T}{|\mathcal{N}|-1}.
	\end{align}
	Combining (i) and (ii), \eqref{eq-optimalB} follows.
\end{proof}
It is observed from Theorem 2 that for any given $p$ satisfying $0\le p<c$, the optimal $B^*$ is unique. Besides, as pointed out in \eqref{Eq-HomoMNO1},  $p^*$ is bounded by $c$. Thus, the optimal $p^*$ can be obtained by searching over the region $[0,c]$. Therefore, the Stackelberg game is solved, and the SE always exists since there exists a unique subgame NE for any given $p$ and $B$.

\section{The Salary-Only Scheme}\label{sec-salaryonly}
In this section, we propose a simple incentive mechanism referred to as salary only scheme. In this scheme, the MNO  motivates WiFi APs to offload data by paying only salary.  Thus, compared with the salary-plus-bonus scheme, there is no bonus part in this scheme. The game formulation and its optimal solutions are given in the following two subsections.  

\subsection{The Game at WiFi APs:}
By removing the bonus part in Problem 1, we can easily get the problem formulation for the game at WiFi APs under the salary only scheme, which is

\underline{\textbf{Problem 3:}}
\begin{align}
\max_{d_i\ge0}~~&pd_i-c_id_i,\\
\mbox{s.t.}~~&d_i\le T_i,\label{eq-con}
\end{align}

To find the optimal strategy of WiFi APs, we first look at the best response of each WiFi AP given the MNO's pricing strategy, i.e., given $p$. 

Problem 3 is easy to solve under a given $p$, and its optimal solution is summarized as follows. The best response function of $AP_i$ is
\begin{align}\label{Sal_AP_Op}
d_i^*=\left\{\kern-1mm\begin{array}{ll}
0, &\mbox{if}~p<c_i, \\
T_i, &\mbox{if}~p\ge c_i.
\end{array}
\right.
\end{align}
It is observed that $AP_i$'s strategy does not depend on other APs' strategy, which means that no non-cooperative game happens on the APs' side. Thus, subgame NE analysis is not necessary for this scheme. 

\subsection{The Game at the MNO:}
By removing the bonus part in Problem 2, the problem formulation for the game at the MNO under the salary only scheme can be easily obtained as

\underline{\textbf{Problem 4:}}
\begin{align}
\max_{p}~~&\lambda\ln \left(1+\sum_{i \in \mathcal
	{N}}d_i(p)\right)-p\sum_{i \in \mathcal {N}}d_i(p),\\
\mbox{s.t.}~~&~p\ge 0.
\end{align}

Given APs' strategy, we now study the optimal pricing strategy of the MNO. To find the optimal strategy of the MNO, we need to substitute APs' strategy into Problem 4. 

For the convenience of expression, we introduce an indicator function, which is 
\begin{align}
\chi_i(p)=\left\{\kern-1mm\begin{array}{ll}
0, &\mbox{if}~p<c_i, \\
1, &\mbox{if}~p\ge c_i.
\end{array}
\right.
\end{align}
Then, the best response of $AP_i$ in \eqref{Sal_AP_Op} can be rewritten as 
\begin{align}\label{Sal_AP_Op2}
d_i^*=\chi_i(p)T_i.
\end{align}
Substituting \eqref{Sal_AP_Op2} into Problem 4, we have 

\underline{\textbf{Problem 4a:}}
\begin{align}
\max_{p}~~&\lambda\ln \left(1+\sum_{i \in \mathcal
	{N}}\chi_i(p)T_i\right)-p\sum_{i \in \mathcal {N}}\chi_i(p)T_i,\\
\mbox{s.t.}~~&~p\ge 0.
\end{align}

To solve Problem 4a, we first present the following proposition. 

\underline{\textbf{Proposition 1:} } The optimal $p^*$ for Problem 4a can only take a value from the set $\{c_0, c_1, c_2, \cdots, c_N\}$, where $c_0=0$.
\begin{proof}
	For the convenience of expression, we assume that $c_0<c_1<c_2<\cdots<c_N$.
	
	First, we show that $p^*$ cannot take any value larger than $c_N$. This can be proved by contradiction. Suppose $p^*=p^\prime$, where $p^\prime>c_N$.  Then, the objective function becomes $\lambda\ln \left(1+\sum_{i \in \mathcal
		{N}}T_i\right)-p^\prime\sum_{i \in \mathcal {N}}T_i$, which is smaller than $\lambda\ln \left(1+\sum_{i \in \mathcal{N}}T_i\right)-c_N\sum_{i \in \mathcal {N}}T_i$. This contradicts with our presumption that
	$p^\prime$ is the optimal solution. Thus, it is concluded that  $p^*$ cannot take any value larger than $c_N$. 
	
	Next, we prove that $p^*$ can not take values between any two consecutive  cost coefficient.  This can be proved by contradiction. Suppose $p^*=p^\prime$, where $c_k<p^\prime<c_{k+1}$. Then, the objective function becomes $\lambda\ln \left(1+\sum_{i=1}^kT_i\right)-p^\prime\sum_{i=1}^kT_i$, which is smaller than $\lambda\ln \left(1+\sum_{i=1}^kT_i\right)-c_k\sum_{i=1}^kT_i$. This contradicts with our presumption that
	$p^\prime$ is the optimal solution. The above proof holds for any $k \in \left\{0, 1, \cdots, N-1\right\}$. Thus, it is concluded that  $p^*$ cannot take values between any two consecutive  cost coefficient.  
\end{proof}

Based the results given in Proposition 1, we propose the following algorithm to solve Problem 4a.  

\begin{figure}
	\centering
	\includegraphics*[width=9.5cm]{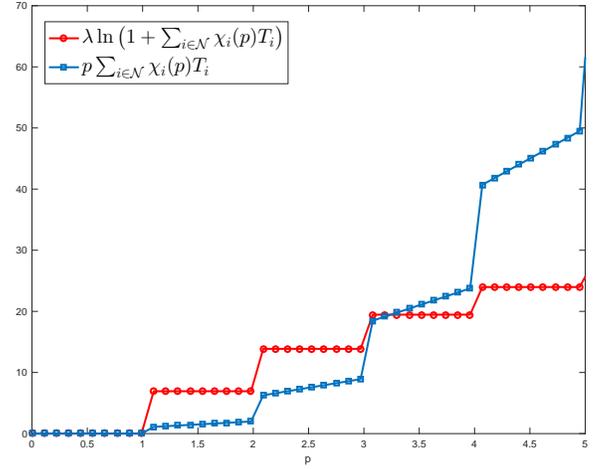}
	\caption{Illustration of Problem 4a's objective function }
	\label{fig:salillufig1}
\end{figure}

\begin{algorithm}
	\caption{Computation of the Optimal Price $p^*$ }
	\begin{algorithmic}[1]
		\State Sort APs in the following order:
		$c(1)<c(2)<\cdots<c(N)$;
		\State Initiate $p^*=0$ \mbox{and} $f^*=0$;
		\For{$i=1; i\le|\mathcal{N}|;i++$}
		\State $p=c(i)$;
		\State $f(i)=\lambda\ln \left(1+\sum_{i \in \mathcal
			{N}}\chi_i(p)T_i\right)-p\sum_{i \in \mathcal {N}}\chi_i(p)T_i$;
		\If{$f(i)>f^*$}
		\State $f^*=f(i)$; and $p^*=c(i)$; 
		\ElsIf{$f(i)<0$}
		\State Break;
		\EndIf
		\EndFor
		\State \textbf{Output:} $p^*$ \mbox{and} $f^*$.
	\end{algorithmic}
\end{algorithm}

It is observed that we will stop the algorithm if a $c_k$ results in a negative value for the objective function (Line 8 and 9 in Algorithm 1), any $c_i$  larger than $c_k$ will not be considered for the optimal solution. This can be explained as follows.  The function $\lambda\ln \left(1+\sum_{i \in \mathcal
	{N}}\chi_i(p)T_i\right)$ is a multi-step increasing function with respect to $p$, and it is capped by $\lambda\ln \left(1+\sum_{i \in \mathcal
	{N}}T_i\right)$.  On the other hand, $p\sum_{i \in \mathcal {N}}\chi_i(p)T_i$ is a piece-wise increasing function with respect to $p$ without a cap. For the convenience of explanation, we plot out these two functions in Fig. \ref{fig:salillufig1}. Thus, after the non-zero  intersection, these two curves diverge, which means the value of the objective function becomes negative forever.  Thus, it is easy to draw the conclusion that the optimal $p^*$ should not lie in the range  after the intersection.

\section{The Bonus-Only Scheme}\label{sec-bonusonly}
In this section, we propose another simple incentive mechanism referred to as bonus only scheme. In this scheme, the MNO motivates WiFi APs to offload data by paying bonus only.  Thus, compared with the salary-plus-bonus scheme, there is no salary part in this scheme. The game formulation and its optimal solutions are given in the following two subsections.  

\subsection{The Game at WiFi APs:}
By removing the salary part in Problem 1, we can easily get the problem formulation for the game at WiFi APs under the bonus only scheme. However, the way to solve the problem is exactly the same as that used to solve Problem 1.  Thus, to make the problem more interesting and mathematically tractable, we introduce a penalty item $\lambda_i (d_i-T_i)$ to the objective function to remove the constraint $d_i\le T_i$, where $\lambda_i$ is the penalty coefficient for $AP_i$. Then, the resultant problem formulation ca be written  as follows.  

\underline{\textbf{Problem 5:}}
\begin{align}
\max_{d_i}~~ &\frac{w_id_i}{\sum_{j \in \mathcal {N}}w_jd_j}B-c_id_i-\lambda_i\left(d_i-T_i\right),\\
\mbox{s.t.}~~& d_i \ge 0.
\end{align}

It can be seen that the penalty item $\lambda_i (d_i-T_i)$ is positive when $d_i>T_i$, which will decrease the value of the objective function of $AP_i$. This indicates the item $\lambda_i (d_i-T_i)$ punish $AP_i$ if it offloads more data than $T_i$.  Thus, it can be seen that the penalty item plays a similar role as the constraint $d_i\le T_i$. Both of them encourage APs not to offload more data than $T_i$. 

Now, we show how to solve this problem. The objective function of Problem 5 is concave in $d_i$, and the constraint is linear. Thus, Problem 5 is a convex optimization problem. Thus, Problem 5 can be solved in the following way. 

First, we take the derivative of $\mathcal {U}_i^F$ with respect to $ d_i$, which leads to
\begin{align}
\frac{\partial\mathcal {U}_i^F}{\partial d_i}=\frac{Bw_i\sum_{j
		\in \mathcal {N}/\{i\}} w_jd_j}{\left(\sum_{j \in \mathcal {N}}
	w_jd_j\right)^2}-\left(c_i+\lambda_i\right),
\end{align}

Next, we let $\frac{\partial\mathcal {U}_i^F}{\partial
	d_i}\big|_{d_i=d_i^\circ}=0$, which leads to
\begin{align}\label{P5_dDericative}
d_i^\circ=\sqrt{\frac{B\sum_{j \in \mathcal {N}/\{i\}}
		w_jd_j}{w_i(c_i+\lambda_i)}}-\frac{\sum_{j \in \mathcal {N}/\{i\}}
	w_jd_j}{w_i}.
\end{align}

Then, due to the fact that $d_i \ge 0$, the optimal solution of Problem 5 can be summarized in the following theorem.

\underline{\textbf{Theorem 3}:} The best response function of $\mbox{AP}_i$ is
\begin{align}\label{bestdforPP5}
d_i^*=\left\{\kern-1mm\begin{array}{ll}
0, &\mbox{if}~d_i^\circ<0, \\
d_i^\circ, &\mbox{if}~d_i^\circ\ge 0,
\end{array}
\right.
\end{align}
where $d_i^{\circ}$ is given in \eqref{P5_dDericative}. 

Now, we investigate the NE of this subgame. From the best response function given in \eqref{bestdforPP5}, it can be observed that WiFi APs can be divided into two categories at the NE.  One category of WiFi APs will be inactive, i.e., not participate in the data offloading. The other category of WiFi APs will  be active, i.e., offloading data at $d_i^\circ$. For the convenience of analysis, we use $\mathcal{S}$ to denote the set of WiFi APs that are active at the NE.

\underline{\textbf{Proposition 2:} } At the NE, there are at least two active WiFi APs, i.e., $|\mathcal{S}|\ge2$.
\begin{proof}
	Proposition 2 can be proved by contradiction. 
	
	First, we suppose $|\mathcal{S}|=0$ at the NE. This means $d_i^{ne}=0,\forall i \in \mathcal{N}$, and all APs' utilities are zero. It is easy to observe that AP $j$ can increase its utility from zero to a positive value by unilaterally changing its $d_j^{ne}$ from 0 to any value in the range $\left(0,\frac{B+\lambda_jT_j}{c_j+\lambda_j}\right)$. This contradicts with the NE assumption. Thus, $|\mathcal{S}|\neq 0$. 
	
	Next, we suppose $|\mathcal{S}|=1$ at the NE. This indicates there is an active AP at the NE, and we denote that AP by $j$. Thus, it follows $d_j^{ne}>0$, and $d_i^{ne}=0, \forall i \in \mathcal{N}/\{j\}$. The utility of $j$ at the NE is $\mathcal{U}^F_j=B-c_jd_j^{ne}-\lambda_jd_j^{ne}+\lambda_jT_j>0$. It is easy to observe that AP $j$ can increase its utility by unilaterally changing $d_j^{ne}$ to a smaller positive value.  This contradicts with the NE assumption. Thus, $|\mathcal{S}|\neq 1$. 
	
	Combining the above results, we conclude that $|\mathcal{S}|\ge2$, i.e., 
	there are at least two active APs at the NE.
\end{proof}

\underline{\textbf{Theorem 4}:} Let $\mathcal{S}$ denote the set of WiFi APs that are active at the NE, the optimal strategy of each AP at the NE is then given as follows.
\begin{align}\label{eq-dioptimal2}
d_i^{ne}\kern-1mm=\kern-1mm\left\{\kern-1mm\begin{array}{ll}\frac{B\left(|\mathcal {S}|-1\right)}{w_i\sum_{j \in \mathcal {S}}\frac{c_j+\lambda_j}{w_j}}\left(1-\frac{c_i+\lambda_i}{w_i}\frac{\left(|\mathcal {S}|-1\right)}{\sum_{j \in \mathcal {S}}\frac{c_j+\lambda_j}{w_j}}\right), & \mbox{if} ~i \in \mathcal{S},\\
0,&\mbox{if} ~i \notin \mathcal{S}.
\end{array}\right.
\end{align}	

\begin{proof}
	From the best response function given in \eqref{bestdforPP5} , it is easy to observe that $d_i^{ne}=0, \forall i \notin \mathcal{S}$. 
	
	Now, we look at the APs that are active at the NE.	
	For any $i \in \mathcal{S}$, it follows from \eqref{P5_dDericative} that  
	\begin{align}
	d_i^{ne}=\sqrt{\frac{B\sum_{j \in \mathcal {S}/\{i\}}
			w_jd_j^{ne}}{w_i(c_i+\lambda_i)}}-\frac{\sum_{j \in \mathcal {S}/\{i\}}
		w_jd_j^{ne}}{w_i},
	\end{align}	  
	which be rewritten as 
	\begin{align}
	\sum_{j \in \mathcal {S}}
	w_jd_j^{ne}=\sqrt{\frac{w_iB\sum_{j \in \mathcal {S}/\{i\}}
			w_jd_j^{ne}}{(c_i+\lambda_i)}}.
	\end{align}	  
	Then, it follows
	\begin{align}\label{eq-dioptimal}
	\frac{c_i+\lambda_i}{w_iB}\left(\sum_{j \in \mathcal {S}}
	w_jd_j^{ne}\right)^2=\sum_{j \in \mathcal {S}/\{i\}}w_jd_j^{ne}.
	\end{align}	 
	For the convenience of expression, we label WiFi APs in $\mathcal{S}$ at the NE
	as $\left\{1, 2, \cdots, |\mathcal{S}|\right\}$.  Then, we have 
	\begin{align}
	\frac{c_1+\lambda_1}{w_1B}\left(\sum_{j \in \mathcal {S}}
	w_jd_j^{ne}\right)^2&=\sum_{j \in \mathcal {S}/\{1\}}w_jd_j^{ne},\label{eq-1}\\
	&~~\vdots\\
	\frac{c_{|\mathcal{S}|}+\lambda_{|\mathcal{S}|}}{w_{|\mathcal{S}|}B}\left(\sum_{j \in \mathcal {S}}
	w_jd_j^{ne}\right)^2&=\sum_{j \in \mathcal {S}/\{|\mathcal{S}|\}}w_jd_j^{ne}\label{eq-S},
	\end{align}	 
	Summing up equations from \eqref{eq-1} to \eqref{eq-S}, we have 
	\begin{align}
	\left(\frac{c_1+\lambda_1}{w_1}+\cdots+\frac{c_{|\mathcal{S}|}+\lambda_{|\mathcal{S}|}}{w_{|\mathcal{S}|}}\right)\frac{\left(\sum_{j \in \mathcal {S}}
		w_jd_j^{ne}\right)^2}{B}\nonumber\\=\left(|\mathcal {S}|-1\right)\sum_{j \in \mathcal {S}}w_jd_j^{ne}.
	\end{align}	 
	Then, it follows
	\begin{align}\label{eq-sumwd}
	\sum_{j \in \mathcal {S}}
	w_jd_j^{ne}=\frac{B\left(|\mathcal {S}|-1\right)}{\sum_{j \in \mathcal {S}}\frac{c_j+\lambda_j}{w_j}}.
	\end{align}	
	Subsitituting \eqref{eq-sumwd} into \eqref{eq-dioptimal}, we have
	\begin{align}
	\frac{c_i+\lambda_i}{w_iB}\left(\frac{B\left(|\mathcal {S}|-1\right)}{\sum_{j \in \mathcal {S}}\frac{c_j+\lambda_j}{w_j}}\right)^2=\frac{B\left(|\mathcal {S}|-1\right)}{\sum_{j \in \mathcal {S}}\frac{c_j+\lambda_j}{w_j}}-w_id_i^{ne},
	\end{align}	 
	which can be rewirtten as
	\begin{align}
	d_i^{ne}=\frac{B\left(|\mathcal {S}|-1\right)}{w_i\sum_{j \in \mathcal {S}}\frac{c_j+\lambda_j}{w_j}}\left(1-\frac{c_i+\lambda_i}{w_i}\frac{\left(|\mathcal {S}|-1\right)}{\sum_{j \in \mathcal {S}}\frac{c_j+\lambda_j}{w_j}}\right).
	\end{align}	
	This finishes the proof of Theorem 4.
\end{proof}

Theorem 4 tells us the best strategies of WiFi APs at the NE, but does not tell us which WiFi APs will be in $\mathcal{S}$. To find which WiFi APs are active at the NE, we first present the following propositions. 

\underline{\textbf{Proposition 3:} }A WiFi AP $i$ is active at the NE if and only if the following condition holds:
\begin{align}\label{eq-dioptimal3}
\frac{c_i+\lambda_i}{w_i}<\frac{\sum_{j \in \mathcal {S}}\frac{c_j+\lambda_j}{w_j}}{|\mathcal {S}|-1}.
\end{align}
\begin{proof}
	This proof consists of two parts: the necessity proof and the sufficiency proof, which are given as follows.
	
	\textbf{Part I: Necessity.} From \eqref{bestdforPP5}, we know all APs in $\mathcal{S}$ must satisfy $d_i^{ne}>0$. Then, it follows from \eqref{eq-dioptimal2} that 
	\begin{align}
	1-\frac{(c_i+\lambda_i)}{w_i}\frac{\left(|\mathcal {S}|-1\right)}{\sum_{j \in \mathcal {S}}\frac{c_j+\lambda_j}{w_j}}>0, \forall i \in \mathcal{S},
	\end{align}
	which can be rewritten as 
	\begin{align}
	\frac{c_i+\lambda_i}{w_i}<\frac{\sum_{j \in \mathcal {S}}\frac{c_j+\lambda_j}{w_j}}{|\mathcal {S}|-1}, \forall i \in \mathcal{S}.
	\end{align}	
	Thus, it is clear that for any AP $i$ that is active at the NE, the condition given in \eqref{eq-dioptimal3} holds. 
	
	\textbf{Part II: Sufficiency.} This part can be proved by contradiction. Suppose that AP $i$ satisfies \eqref{eq-dioptimal3} but is not active at the NE, i.e., $i \notin \mathcal{S}$, which means $d_i^{ne}=0$.
	
	Let us look at the derivative of $\mathcal {U}_i^F$ at the NE, which is 
	\begin{align}
	\frac{\partial\mathcal {U}_i^F}{\partial d_i}\big|_{d_i=d_i^{ne}}&=\frac{Bw_i\sum_{j
			\in \mathcal {N}/\{i\}} w_jd_j^{ne}}{\left(\sum_{j \in \mathcal {N}}
		w_jd_j^{ne}\right)^2}-\left(c_i+\lambda_i\right)\nonumber\\
	&\overset{a}{=}\frac{Bw_i\sum_{j
			\in \mathcal {S}} w_jd_j^{ne}}{\left(\sum_{j \in \mathcal {S}}
		w_jd_j^{ne}\right)^2}-\left(c_i+\lambda_i\right)\nonumber\\
	&=\frac{Bw_i}{\sum_{j \in \mathcal {S}}
		w_jd_j^{ne}}-\left(c_i+\lambda_i\right)\nonumber\\
	&\overset{b}{=}\frac{w_i\sum_{j \in \mathcal {S}}\frac{c_j+\lambda_j}{w_j}}{|\mathcal {S}|-1}-\left(c_i+\lambda_i\right),
	\end{align}
	where $''a''$ results from the fact that $d_j^{ne}=0, \forall j \notin \mathcal{S}$, and $''b''$ results from \eqref{eq-sumwd}.
	
	Then, it follows 
	\begin{align}\label{eq-derivativelargerthan}
	\frac{\partial\mathcal {U}_i^F}{\partial d_i}\big|_{d_i=d_i^{ne}}=\frac{1}{w_i}\left(\frac{\sum_{j \in \mathcal {S}}\frac{c_j+\lambda_j}{w_j}}{|\mathcal {S}|-1}-\frac{c_i+\lambda_i}{w_i}\right)>0,
	\end{align}
	where the inequality results from the presumption that AP $i$ satisfies \eqref{eq-dioptimal3}.
	
	The inequality \eqref{eq-derivativelargerthan} indicates that AP $i$ can increase it utility by unilaterally increasing its data offloading amount from zero. This contradicts with the presumption of NE. 
	Thus, it is concluded that for any AP $i$ satisfying  \eqref{eq-dioptimal3},  it must be active at the NE.
	
	Combine the results obtained in Part I and II,  we conclude that an AP $i$ is active at the NE if and only if $
	\frac{c_i+\lambda_i}{w_i}<\frac{\sum_{j \in \mathcal {S}}\frac{c_j+\lambda_j}{w_j}}{|\mathcal {S}|-1}$ holds. This proves Proposition 3. 
\end{proof}

\underline{\textbf{Proposition 4: } }For any two APs $m$ and $n$ satisfying the condition  $\frac{c_m+\lambda_m}{w_m}<\frac{c_n+\lambda_n}{w_n}$, if $n$ is active at the NE,  then $m$ must also be active at the NE. 

\begin{proof}
	This can be proved by contradiction. Let us make the assumption that $m$ is not active at the NE, which indicates \eqref{eq-dioptimal3} does not hold for $m$. 
	
	Let $\mathcal{R}$ denote the set of APs that are active at the NE except $n$, and suppose that $n$ is active at the NE.  Then, we have $\mathcal{S}=\mathcal{R}\bigcup\{n\}$, and it follows from 
	\eqref{eq-dioptimal3} that 
	\begin{align}
	\frac{c_n+\lambda_n}{w_n}<\frac{\sum_{j \in \mathcal {R}}\frac{c_j+\lambda_j}{w_j}+\frac{c_n+\lambda_n}{w_n}}{|\mathcal {S}|-1},
	\end{align}
	which can be rewritten as
	\begin{align}
	\left(|\mathcal {S}|-2\right)\frac{c_n+\lambda_n}{w_n}<\sum_{j \in \mathcal {R}}\frac{c_j+\lambda_j}{w_j}.
	\end{align}
	Since $\frac{c_m+\lambda_m}{w_m}<\frac{c_n+\lambda_n}{w_n}$, it follows 
	\begin{align}
	\left(|\mathcal {S}|-2\right)\frac{c_m+\lambda_m}{w_m}<\sum_{j \in \mathcal {R}}\frac{c_j+\lambda_j}{w_j},
	\end{align}
	which can be rewritten as
	\begin{align}\label{eq-mineq}
	\frac{c_m+\lambda_m}{w_m}<\frac{\sum_{j \in \mathcal {R}}\frac{c_j+\lambda_j}{w_j}+\frac{c_m+\lambda_m}{w_m}}{|\mathcal {S}|-1}.
	\end{align}
	According to Proposition X, \eqref{eq-mineq} indicates that $m$ should be active at the NE.
	This contradicts with the presumption that $m$ is not active at the NE. 
	
	Thus, it is concluded $m$ must be active at the NE. This proves Proposition 4.
\end{proof}

Proposition 4 tells us that we should always include the AP with smaller $\frac{c_i+\lambda_i}{w_i}$ into $\mathcal{S}$ first. Proposition 3 tells us that we should only include APs satisfying \eqref{eq-dioptimal3} into $\mathcal{S}$. Thus, based on these results, we summarize the method to compute the subgame NE in the following table.

\begin{algorithm}
	\caption{Computation of the Subgame NE}
	\begin{algorithmic}[1]
		\State Sort APs in the following order:
		$\frac{c_1+\lambda_1}{w_1}<\frac{c_2+\lambda_2}{w_2}<\cdots<\frac{c_{|\mathcal{N}|}+\lambda_{|\mathcal{N}|}}{w_{|\mathcal{N}|}}$;
		\State Initiate the set $\mathcal{S}$ by letting  $\mathcal{S}=\left\{AP_1, AP_2\right\}$ ;
		\State Initiate $i$ by setting $i=3$;
		\While{$i\le|\mathcal{N}|$ \textbf{and} 	$\frac{c_i+\lambda_i}{w_i}<\frac{\sum_{j \in \mathcal {S}}\frac{c_j+\lambda_j}{w_j}+\frac{c_i+\lambda_i}{w_i}}{|\mathcal {S}|}$} 
		\State Put $AP_i $ into $\mathcal{S}$;
		\State Update $i=i+1$;
		\EndWhile
		\For{$i=1; i\le|\mathcal{S}|;i++$}
		
		\State $d_i^{ne}=\frac{B\left(|\mathcal {S}|-1\right)}{w_i\sum_{j \in \mathcal {S}}\frac{c_j+\lambda_j}{w_j}}\left(1-\frac{c_i+\lambda_i}{w_i}\frac{\left(|\mathcal {S}|-1\right)}{\sum_{j \in \mathcal {S}}\frac{c_j+\lambda_j}{w_j}}\right)$;
		\EndFor
		\State  Set $d_i^{ne}=0, \forall i \in \mathcal{N}/\mathcal{S}$;
		\State \textbf{Output:} $\mbox{NE}=\left(d_1^{ne}, d_2^{ne}, \cdots, d_{|\mathcal{N}|}^{ne}\right)$.
	\end{algorithmic}
\end{algorithm}

\textbf{\textit{Remark:}} It is interesting to observe from Algorithm 2 that the number of active APs at the NE does not depend on the amount of bonus. Whether a WiFi AP is active at the NE only depends on how its  cost to quality ratio  ($\frac{c_i+\lambda_i}{w_i}$) compares with others'. This indicates that the bonus-only scheme cannot increase  the number of WiFi APs participating in the offloading by increasing the bonus. However, WiFi APs selected by  the bonus-only scheme  are premium APs providing high-quality offloading service at low cost. 

\subsection{The Game at the MNO}
By removing the salary part in Problem 2, the problem formulation for the game at the MNO under the bonus only scheme can be easily obtained as

\underline{\textbf{Problem 6:}}
\begin{align}
\max_{B}~~&\lambda\ln \left(1+\sum_{i \in \mathcal
	{N}}d_i(B)\right)-B,\\
\mbox{s.t.}~~&~B\ge 0.
\end{align}

Given the APs' strategies, we now study the best strategy of the MNO. To find the optimal strategy of the MNO, we need to substitute the subgame NE (i.e. the optimal solution of Problem 5)  into the objective funcion of Problem 6, which leads to

\underline{\textbf{Problem 6a:}}
\begin{align}
\max_{B}~~&\lambda\ln \left(1+\sum_{i \in \mathcal
	{S}}H_iB\right)-B,\\
\mbox{s.t.}~~&~B\ge 0.
\end{align}
where $H_i\triangleq\frac{|\mathcal {S}|-1}{w_i\sum_{j \in \mathcal {S}}\frac{c_j+\lambda_j}{w_j}}\left(1-\frac{(c_i+\lambda_i)}{w_i}\frac{\left(|\mathcal {S}|-1\right)}{\sum_{j \in \mathcal {S}}\frac{c_j+\lambda_j}{w_j}}\right).
$

The optimal solution of Problem 6a is given as follows. 

\underline{\textbf{Theorem 5}:} The optimal solution of Problem 6a is
\begin{align}
B^*=\left(\lambda-\frac{1}{\sum_{i\in\mathcal{S}}H_i}\right)^+,
\end{align}
where $(\cdot)^+$ denotes $\max(0,\cdot)$.
\begin{proof}
	The second order derivative of Problem 6a's objective function can be obtained as  $-\frac{\lambda \sum_{i\in\mathcal{S}}H_i^2}{\left(1+B\sum_{i\in\mathcal{S}}H_i\right)^2}$, which is less than zero. Thus, it is concluded that Problem 6a's objective function is concave in $B$. Then, its maximum value can be obtained by setting the first order derivative $\frac{\lambda\sum_{i\in\mathcal{S}}H_i}{\sum_{i\in\mathcal{S}}H_iB+1}-1$ to zero, which results in $B=\frac{\lambda\sum_{i\in\mathcal{S}}H_i-1}{\sum_{i\in\mathcal{S}}H_i}. $ Then, due to the fact that $B \ge 0$, the optimal solution of Problem 6a can be obtained as $B^*=\left(\frac{\lambda\sum_{i\in\mathcal{S}}H_i-1}{\sum_{i\in\mathcal{S}}H_i}\right)^+.$ 
\end{proof}

\section{Numerical Results}\label{sec-numbericalresults}
In this section, numerical examples are given to investigate the performance of the proposed incentive mechanisms.
\subsection{Study on the Salary-Plus-Bonus Scheme}
\subsubsection{Subgame NE Analysis}
In this numerical example, we assume that there are three heterogeneous APs in the HetNet. The simulation parameters are chosen as follows: $c_1=2$, $c_2=3$, $c_3=4$; $T_1=T_2=T_3=5$; $w_1=0.2$, $w_2=0.3$, $w_3=0.5$. We investigate the subgame  equilibrium behavior of WiFi APs for given $p$ and $B$. It is observed that for the same $p$ and $B$, the AP with lower cost is willing to offload more data. Increasing the salary rate $p$ is more effective in boosting up the amount of data offloaded. The AP with lower $w_i$ is more sensitive to the bonus changes. When the bonus increases, the amount of offloaded data increases faster for the AP with lower $w_i$. This is because that the bonus distribution not only depends on $d_i$ but also depends on $w_i$ in the proposed incentive mechanism. Thus, in order to get the same partition of the bonus, the AP with lower $w_i$ must offload more data than the AP with higher $w_i$. This also indicates that the proposed bonus scheme is bias towards to the AP with good quality of service $w_i$, and thus provides an incentive for APs to improve their service quality.
\begin{figure}[t]
	\centering
	\includegraphics*[width=9cm]{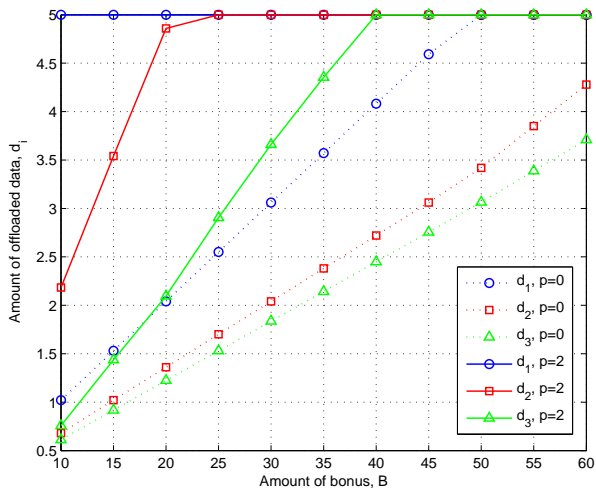}
	\caption{Subgame nash equilibrium of three heterogeneous WiFi APs under different bonus $B$ and different salary rate $p$.}
	\label{Fig2}
\end{figure}

\subsubsection{The Utility of the MNO}
\begin{figure}[t]
        \centering
        \includegraphics*[width=9cm]{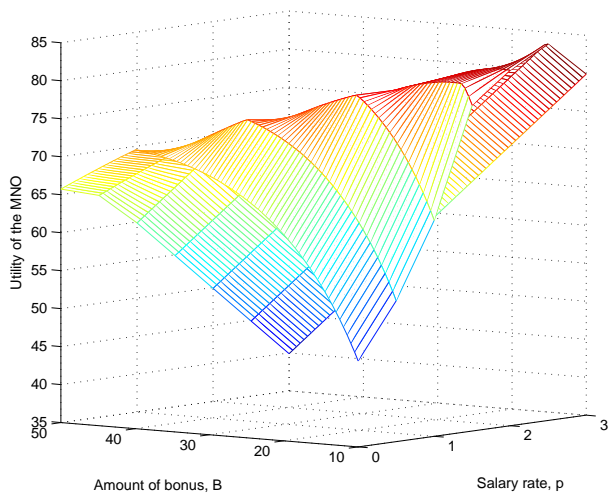}
        \caption{Utility of the MNO under the Stackelberg game formulation.}
        \label{Fig1}
\end{figure}

In this numerical example, we assume that there are two heterogeneous APs existing in the HetNet. The simulation parameters are given as $c_1=2$, $c_2=3$, $T_1=T_2=5$, $w_1=0.2$, $w_2=0.3$, $\lambda=50$. It is observed from Fig. \ref{Fig1} that the utility function of the MNO is neither convex nor concave in $p$ and $B$. It is also observed that when both $p$ and $B$ are large or small, the utility of the MNO is low. While the MNO's utility is high when one of them (either $p$ or $B$) is large and the other one of them is small. This indicates that the MNO should in general adopt either the low-salary high-bonus strategy or the high-salary low-bonus strategy to achieve high utilities.

\subsection{Comparison Among Three Incentive Mechanisms}
\begin{figure}[t]
	\centering
	\includegraphics*[width=9.5cm]{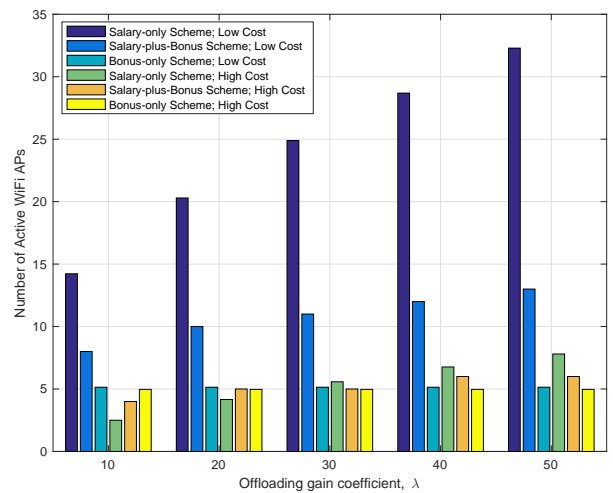}
	\caption{Comparison of the number of active WiFi APs.}
	\label{Fig3}
\end{figure}

\begin{figure}[t]
	\centering
	\includegraphics*[width=9.5cm]{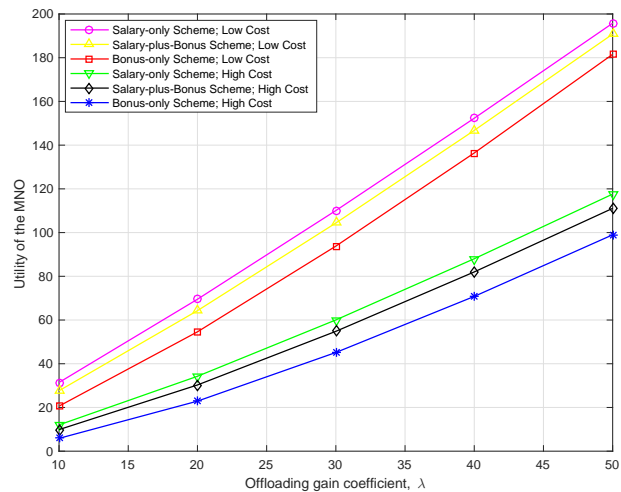}
	\caption{Comparison of the utility of the MNO.}
	\label{Fig4}
\end{figure}

For the purpose of comparison, we use the same simulation setup for three incentive mechanisms. The simulation setup is as follows. We totally generate $100$ WiFi APs. The offloading limits of each WiFi AP is uniformly drawn from the range  $(0, 5]$. The offloading quality of each WiFi AP is uniformly drawn from the range  $(0, 1]$.  For the bonus-only scheme, the penalty factor $\lambda_i$ for each WiFi AP is set equal to the inverse of its offloading limit $T_i$. For the purpose of studying the effect of WiFi APs' cost on the proposed schemes, we consider two sets of cost values, i.e. the low cost set and the high cost set.   For the lost cost set, the cost of  each WiFi AP is uniformly drawn from the range  $(0, 1]$. While for the high cost set, the cost of each WiFi AP is uniformly drawn from the range $[1,10]$.  The results given in Fig. \ref{Fig3} ad \ref{Fig4} are obtained by averaging over $1000$ simulation runs. 

\subsubsection{Suboptimal Solution for the Salary-Plus-Bonus Scheme} Though the simplicial method in \cite{PHerings} can be applied to find the subgame NE, it  does not apply to games with large-size. Thus, for the purpose of performance comparison for a game with large size, based on the results obtained for the salary-only scheme and the bonus-only scheme,  we develop a suboptimal algorithm (Algorithm 3) to quickly find the subgame NE for the salary-plus-bonus scheme. Substituting the subgame NE found by Algorithm 3 into the Problem 2, it is easy to show the problem is concave in $B$ for a given $p$. Then, under a given $p$, the optimal $B^*$ can be obtained by setting the first order derivative with respect to $B$ to zero.  Then, we do a linear search for $p^*$ over the range $[c_1, c_N]$.   
\begin{algorithm}
	\caption{A suboptimal strategy of WiFi APs}
	\begin{algorithmic}[1]
		\State Sort APs in the following order:
		$c_1<c_2<\cdots<c_N$.
		\State Put all APs satisfying $c_i<p$ into the set $\mathcal{S}_T$.
		\If{ there is no AP or one AP left (denoted by $l$) in $\mathcal{N}$} 
			\State  Set $d_i^{ne}=T_i, \forall i \in \mathcal{S}_T$;
		\State  Set $d_l^{ne}=\sqrt{\frac{B\sum_{j \in \mathcal {S}_T} w_jT_j}{a_lw_l}}-\frac{\sum_{j \in \mathcal {S}_T} w_jT_j}{w_l}$;
		\Else
		\State Sort the remaining APs in the following order: 
		
		$\frac{a_1}{w_1}<\frac{a_2}{w_2}<\cdots<\frac{a_{|\mathcal{N}|}}{w_{|\mathcal{N}|}}$;
		\State Initiate the set $\mathcal{S}$ by letting  $\mathcal{S}=\left\{AP_1, AP_2\right\}$ ;
		\State Initiate $i$ by setting $i=3$;
		\While{$i\le|\mathcal{N}|$ \textbf{and} 	$\frac{a_i}{w_i}<\frac{\sum_{j \in \mathcal {S}}\frac{a_j}{w_j}+\frac{a_i}{w_i}}{|\mathcal {S}|}$} 
		\State Put $AP_i $ into $\mathcal{S}$;
		\State Update $i=i+1$;
		\EndWhile
		\For{$i=1; i\le|\mathcal{S}|;i++$}
		
		\State $d_i^{ne}=\frac{B\left(|\mathcal {S}|-1\right)}{w_i\sum_{j \in \mathcal {S}}\frac{a_j}{w_j}}\left(1-\frac{a_i}{w_i}\frac{\left(|\mathcal {S}|-1\right)}{\sum_{j \in \mathcal {S}}\frac{a_j}{w_j}}\right)$;
		\If{$d_i^{ne}>T_i$}
		\State Remove $AP_i$ from $\mathcal{S}$; and Put $AP_i$ into $\mathcal{S}_T$;
		\EndIf
		\EndFor
		\State  Set $d_i^{ne}=T_i, \forall i \in \mathcal{S}_T$;
		\State  Set $d_i^{ne}=0, \forall i \in \mathcal{N}/(\mathcal{S}\bigcup \mathcal{S}_T)$;
			\EndIf
		\State \textbf{Output:} $\left(d_1^{ne}, d_2^{ne}, \cdots, d_{|\mathcal{N}|}^{ne}\right)$.
	\end{algorithmic}
\end{algorithm}

\subsubsection{The Number of Active WiFi APs} The simulation results for this example are shown in Fig. \ref{Fig3}.
Firstly, it is observed that the number of active WiFi APs increases with the increasing of the offloading gain coefficient for the salary-only scheme. This is as expected. A higher offloading gain coefficient means that the MNO benefits more from offloading data. Thus, the MNO is more willing to set up a higher salary rate $p$ for the salary-only scheme. For the salary-only scheme, whether a WiFi AP will join the offloading only depends on the salary rate $p$ and its offloading cost $c$. Thus, for the same set of cost values, a higher $p$ indicates more WiFi APs will join the offloading. 

Secondly, it is interesting to observe that the number active WiFi APs remains the same for the bonus-only scheme no matter how the offloading gain coefficient changes. The reason is as follows. It can be seen from Proposition 3 and Algorithm 2 that whether a WiFi AP is active at the equilibrium or not only depends on the $\frac{c_i+\lambda_i}{w_i}<\left(\sum_{j \in \mathcal {S}}\frac{c_j+\lambda_j}{w_j}+\frac{c_i+\lambda_i}{w_i}\right)\big/|\mathcal {S}|$, which is not related to the offloading gain coefficient.  

Thirdly, it is observed the number of active WiFi APs under the bonus-only scheme is much less than that under the salary-only scheme. This is due to the following reason. For the salary-only scheme,  a WiFi AP will be admitted into the active as long as the salary rate $p$ is larger than its offloading cost $c$. However, for the bonus-only scheme, only when a AP $i$ satisfies the condition $\frac{c_i+\lambda_i}{w_i}<\left(\sum_{j \in \mathcal {S}}\frac{c_j+\lambda_j}{w_j}+\frac{c_i+\lambda_i}{w_i}\right)\big/|\mathcal {S}|$, it will be admitted into the set $|\mathcal {S}|$. Since this condition is more difficult to satisfy, the number of active APs is much less. The good side is that it can help the bonus-only scheme select APs with high offloading quality but low cost (i.e., low $\frac{c_i+\lambda_i}{w_i}$). 

Fourthly, it is observed that the number of active WiFi APs for the high cost set is lower than that of the low cost set for the salary-only scheme. However, the number of active WiFi APs for the high cost set is almost the same as that of the low cost set for the bonus-only scheme. This is due to the fact that for the salary-only scheme, the number of active WiFi APs strongly depends on the cost of the APs. However, for the bonus-only scheme, the number of active WiFi APs depends on the $c_i$, $\lambda_i$, $w_i$, and the relationship among APs.  Thus, the effect of APs' cost on the number of active APs at the NE is much weaker. 

\subsubsection{The Utility of the MNO} The simulation results for the comparison of the MNO's utility under different incentive mechanisms are shown in Fig. \ref{Fig4}.  Firstly, it is observed that the MNO's utility under the salary-only scheme is higher than that under the bonus-only scheme. This is because the bonus-only scheme only is more picky on the WiFi APs. It only selects APs with high quality to cost ratio. Thus, the number of APs working for the MNO is much lower,  which leads to a low utility.  Secondly, for all three schemes, the MNO's utility under the low cost set is larger than that under the high cost set, and the MNO's utility increases with the increasing of the offloading gain coefficient. This is as expected. Look at Fig. \ref{Fig3} and Fig. \ref{Fig4} together, an important finding is that the salary-plus-bonus scheme can achieve almost the same utility as the salary-only scheme, using much less number of active WiFi APs.  This indicates that the salary-plus-bonus is more effective in selecting the good WiFi APs. It not only selects WiFi APs with high quality to cost ratio, but also selects WiFi APs with low cost. In this way, it not only guarantees the offloading quality but also effectively increases the amount of data offloaded. It inherits the advantages from both schemes and strikes a well balance between the offloading quality and the offloading data mount. 

\section{Conclusion}\label{sec-conclusion}
In this paper, we proposed a salary-plus-bonus incentive mechanism to motivate WiFi APs for providing data offloading service to the MNO in a heterogeneous network. Under the proposed salary-plus-bonus scheme, we investigated the interactions between WiFi APs and the MNO using the Stackelberg game. We then studied the formulated Stackelberg game under two different scenarios: Homogeneous APs and Heterogeneous APs. For both cases, we derived the best response functions for WiFi APs (i.e. the optimal amount of data to offload), and showed that the Nash Equilibrium (NE) always exists for the subgame. Then, given WiFi APs' strategies, we investigated the optimal strategy (i.e. the optimal salary and bonus) for the MNO to maximize its utility.  Closed-form solutions have been obtained for the homogeneous  case. For the heterogeneous case, closed-form solutions have been obtained for the two-AP case. We then proposed two simple incentive mechanisms, which are the \emph{salary-only} scheme and the \emph{bonus-only} scheme. For both schemes, we developed low-complexity algorithms to find both WiFi APs' and the MNO's optimal strategy. It have been shown that the salary-only scheme is more effective in motivating more APs to offload more data, while the bonus-only scheme is more effective in selecting premium APs which can provide high-quality offloading service at low cost. Then, for the purpose of investigating the performance of the salary-plus-bonus scheme for heterogeneous networks with  large-size, we  developed a  low-complexity suboptimal algorithm to quickly find the strategy of APs and the MNO.  It have been shown by simulations that the salary-plus-bonus scheme can strike a  good balance between the offloading quality and offloaded data volume.

\end{document}